\documentclass{jpsj2}
\usepackage{graphicx}
\def\runtitle{Giant Quantum Reflection of Neon Atoms}
\def\runauthor{Fujio \textsc{Shimizu} and Jun-ichi \textsc{Fujita}}

\title{%
Giant Quantum Reflection of Neon Atoms from a Ridged Silicon
Surface
}

\author{%
Fujio \textsc{Shimizu} and Jun-ichi \textsc{Fujita}$^{1,}$
}

\inst{%
Institute for Laser Science, University of
Electro-Communications, Chofu-shi, Tokyo 182-8585
\\
$^1$NEC Fundamental Research Laboratories, 34 Miyukigaoka, 
Tsukuba 305-8501
}

\recdate{\today}

\abst{%
The specular reflectivity of slow, metastable neon atoms from a silicon
surface was found to increase markedly when the flat surface was replaced
by a grating structure with parallel narrow ridges.
For a surface with ridges that have a sufficiently narrow top,
the reflectivity was found to increase more than two orders of 
magnitude at the incident angle $\theta$ of
10~mRad from the surface. The slope of the reflectivity vs
$\theta$ near zero was found to be nearly an order of 
magnitude smaller than that of a flat surface. 
A grating with 6.5~\% efficiency for the first-order 
diffraction was fabricated by using the ridged surface structure.
}

\kword{%
quantum, reflection, silicon, surface, neon, laser cooling,
atom optics, grating, beam splitter
}

\begin{document}
\sloppy
\maketitle

Quantum reflection\cite{Shimizu} of a particle is a reflection of
wave nature that occurs as a result of impedance  mismatch at a 
steep slope of interacting potential. The reflection occurs equally
even when the particle is moving along the downhill slope of the potential. 
The reflectivity depends on the steepness of the spatial variation of
the potential and approaches unity when the kinetic energy of the particle
approaches zero. Quantum reflection was
verified experimentally 
by the reflection of helium and hydrogen atoms on a liquid helium 
surface\cite{Nayak,Berkhout,Doyle,Yu} and, more 
recently, by the detection of specular reflection
of metastable neon atoms from silicon and glass
surfaces\cite{Shimizu}, which was caused by the attractive 
van der Waals potential 
near the surface.
The quantum reflection due to the van der Waals potential
occurs at a distance many times larger than the 
atomic separation of the solid, and the atomic scale irregularity of the solid
surface is averaged out.
The surface
of the deflection is clearly defined within the distance of the de Broglie
wavelength of an atom colliding on the surface. 
These features are in
remarkable contrast to those of atom reflectors developed to date and
closely resemble classical optical reflectors. Therefore, in principle,
it should be possible to use quantum 
reflection as a tool to make
simple, stable and nondispersive atom reflectors.
However, the reflectivity from a flat solid surface
decreases so rapidly with atomic 
velocity that its practical application
is almost inconceivable.
We discuss and experimentally demonstrate a new technique that improves
the reflectivity many orders of magnitude by modifying the surface
of the solid.
By making the surface periodically ridged with a narrow flat
top, we improved the reflectivity more than two orders of magnitude
at an incident angle of 10~mRad. As an example of its application, 
we fabricated a reflective
grating that had an efficiency of 6.5~\% for the first-order
diffracted atomic wave.

The atom will feel spatial variation of potential sufficiently
steep when 
the change in the local wavevector $k=\sqrt{k_0^2-2mU/\hbar^2}$ within 
the distance $1/k_0$ becomes larger than $|k_0|$, where $k_0$ is the 
wave vector of the atom at a large distance from the surface.
Therefore,
\[
\phi=\frac{1}{k^2}\frac{dk}{dz}\geq 1.
\]
For a power law potential $U(z)=-C_n/z^n$ with $n>2$, 
$\phi$ takes the maximum value
\begin{equation}
\phi_{max}=\frac{1}{z_{max}k_0}\frac{(n+1)(n-2)^{1/2}}
{3^{3/2}n^{1/2}},
\label{phimax}
\end{equation}
at the distance
\[ 
z_{max}=\left\{\frac{n-2}{(n+1)}\frac{mC_n}{\hbar^2k_0^2}\right\}^{1/n}.
\]
from the surface\cite{Carraro}. 
The atomic wave is partially reflected at the vicinity of
$z_{max}$, and the part of the wave that passes 
through the peak of $\phi$
is accelerated adiabatically and hits the
repulsive wall of the surface.
The $\phi$ increases
indefinitely as the wave vector $k_0$ or the potential constant $C_n$
approaches zero, and the reflectivity
is expected to approach unity if the normal incident velocity
is sufficiently low or $C_n$ is sufficiently small.
Since $C_n$ is approximately proportional to the density of the solid
near the surface, the reflectivity should be improved by decreasing 
the effective density near the surface.
Motivated by this consideration, we etched the silicon surface to form
a grating structure with very narrow ridges so that the atom interacts
with only a small fraction of the entire surface. We measured the
reflectivity of an ultracold neon atomic beam in the
$1s_3$ metastable state as a function of incident angle.

\begin{figure}
\includegraphics[width=0.5\linewidth,height=0.7\linewidth]{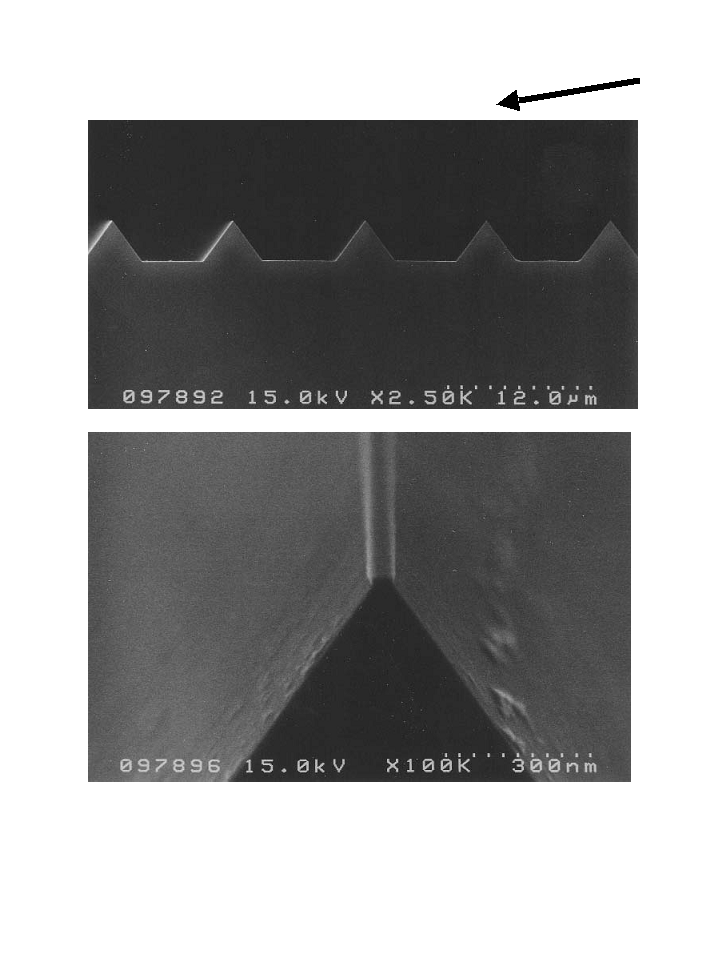}
\caption{Scanning microscope photograph of the silicon grating surface.
Top: cross-sectional view. Bottom: expanded view of the ridge. The grating is
fabricated on the (0,0,1) surface. The ridge runs along the (1,1,0) 
direction. The side walls of the ridge are $(\pm 1,\mp 1,1)$ facets.
The arrow in the top figure indicates the direction of the incident atomic beam.}
\label{ridge}
\end{figure}

An example of the silicon surface used in the experiment 
is shown in Fig.~\ref{ridge}.
The grating structure was formed on the (0,0,1) surface that was 
flat within 20~nm,
by the following procedure. 
A 100~nm-thick oxidized layer was formed on the (0,0,1) 
surface, then the surface
was coated with a negative resist. The pattern of a periodic array of stripes 
that form the top of the ridges was written with an electron beam on the resist,
and the oxidized layer
was removed photo-lithographically with buffered HF
excluding the top of the ridges. 
The direction of the ridge was aligned precisely 
parallel to the (1,1,0) direction.
The silicon was then etched by tetramethylammonium (TMAH). 
The etchant preferentially etched 
the (0,0,1) surface and left $(\pm 1,\mp 1,1)$ facets
on two sides of the oxidized stripe, forming a roof-shaped ridge.
The etching process was stopped just before 
the bottom (0,0,1) facet disappeared,
or the top of the ridge was eroded seriously.
The height of the ridge was between 1.5 and 6~$\mu$m.
Finally, the
oxidized layer on the ridge was removed with buffered HF. The flatness
of the top of the ridge was believed not to be degraded by the etching
process.
The size of the plate was typically 10~mm by 90~mm. The ridge was formed
parallel to the 10~mm side and was perpendicular to the atomic beam.
We measured the reflectivity of several samples with a periodicity 
$p$
between 10 and 100~$\mu$m and a width $d$ of the top of the ridge
between 40~nm and 11$\mu$m.

\begin{figure}
\includegraphics[width=0.6\linewidth,height=0.75\linewidth]{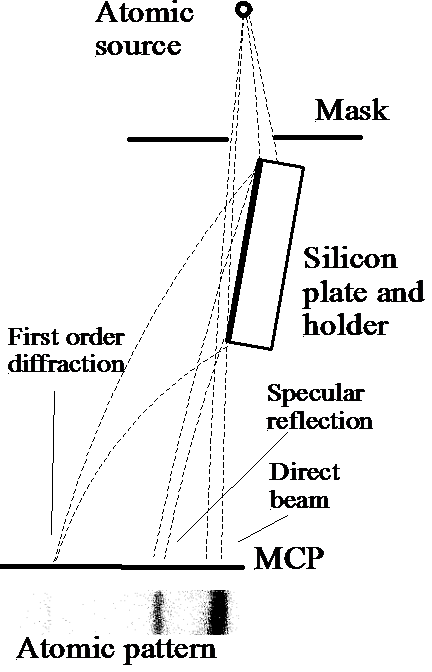}
\caption{Cross-sectional view of the experimental setup.}
\label{experiment}
\end{figure}

The experimental setup was described in our previous report\cite{Shimizu}.
Metastable neon atoms in the $1s_3$ state were generated 
by optical pumping from atoms in the
$1s_5$ state trapped in a magneto-optical trap. The
$1s_3$ atoms 
were freed from the trap and
fell nearly vertically pulled by gravity. A diaphragm with a square hole
($1\times 1$~mm) was placed 33~cm below the atomic source.
The silicon plate was placed 15~cm further below the diaphragm. Its 
surface was placed approximately parallel to a side of the square hole
of the diaphragm. The longitudinal velocity of the atom at the surface
was approximately 3~m/s, and its de Broglie wavelength was 7~nm.
The entire vertical length of the silicon plate was
illuminated by the atoms that passed through the square hole. 
A part of the atomic beam
hit the plate, reflected and formed a line image on the microchannel
plate detector (MCP) that
was placed 112~cm below the atomic source. A part of the atomic beam 
missed the plate and hit the MCP directly. The intensity of the direct
beam was used to calibrate the absolute reflectivity, while the angle
of the reflection was calculated from the distance between the image
of the reflected atoms and that of the direct beam.
Figure~\ref{experiment} shows a cross-sectional view of the experimental
setup and a typical
pattern of atoms on the MCP.
The pattern was obtained from the surface that had 1~$\mu$m wide ridges
and a periodicity of 100~$\mu$m.
The strong stripe in the middle is the image of the specularly reflected
atoms.
The weak line at the left is the first-order diffraction pattern 
produced by the 100~$\mu$m periodic structure.

\begin{figure}
\includegraphics[width=0.7\linewidth,height=0.55\linewidth]{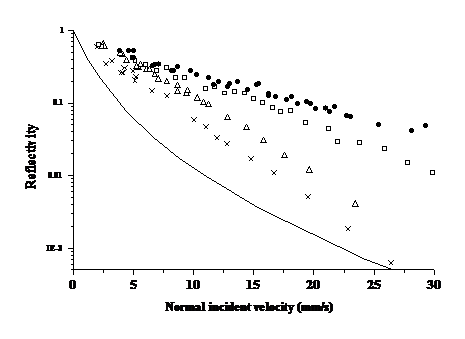}
\caption{The reflectivity as a function of the normal incident velocity
of the atomic beam. The  velocity of
30~mm/s corresponds to the incident angle of 10~mRad. 
The periodicity $p$ and the width of the top
of the ridge $d$ are as follows. X: $p=100$~$\mu$m, $d=11$~$\mu$m;
$\triangle$: $p=100$~$\mu$m, $d=1$~$\mu$m; $\Box$: $p=30$~$\mu$m, 
$d=40$~nm;
$\bullet$: $p=10$~$\mu$m, $d=40$~nm; solid line: flat surface. 
}
\label{reflectivity}
\end{figure}

The reflectivity was measured at the normal incident 
velocity between 4 and
30~mm/s, which corresponded to the incident angle $\theta$ of the atomic beam
measured from the silicon surface between 1.3 and 10~mRad.
Figure~\ref{reflectivity} 
shows the reflectivity of four different
samples as a function of normal incident velocity.
For comparison, the curve for a flat surface taken from
ref.~1 is shown by a solid line.
The periodicity $p$ and the width $d$ of the four samples were
$100$~$\mu$m and $11$~$\mu$m, $100$~$\mu$m and $1$~$\mu$m,
$30$~$\mu$m and $40$~nm, and  $10$~$\mu$m and
$d=40$~nm, respectively.

The one-dimensional wave equation with the potential $-C_n/z^n$ is written
in a dimensionless form if the distance $z$ is normalized by 
$\beta=(mC_n/\hbar)^{1/(n-2)}$ and the energy by $\hbar^2/(m\beta^2)$.
This means that, for the van der Waals potential $-C_3/z^3$, the reflectivity
of a ridged surface is equal to the reflectivity of a flat
surface
at $(d/p)$ times smaller velocity, if
the one-dimensional theory with $C_3$ replaced  by $(d/p)C_3$
is applicable. The actual improvement in the reflectivity is not that 
spectacular and shows that the two-dimensional wave equation must be 
integrated
to obtain a quantitative result. However, the general trend in 
Fig.~\ref{reflectivity} agrees with this estimate.
The reflectivity is larger for a smaller $d/p$. 
The slope near $\theta=0$ for the samples with $d/p \sim 100$
is approximately eight times smaller than that of the flat surface.
As a result,  at around the velocity of 30~mm/s corresponding to
the incident angle $\theta=10$~mRad, the reflectivity was more than 
$10^2$ times larger than that of a flat surface .
For the sample with a larger periodicity, the reflectivity decreased
more rapidly than that of the flat surface at large $\theta$. 
This result is easily
understood based on geometrical consideration.
When an atom passes near the top of the ridge, 
it must pass within distance 
$z_{max}$  in order 
to be deflected and to contribute to the specular
reflection.
An atom that passes at a larger distance can be deflected at the next
ridge only when it does not hit the side of the ridge.
By ignoring the numerical factor and putting $\phi_{max}\sim 1$ in 
Eq.~(\ref{phimax}), $z_{max}$ is roughly given by 
$1/k_0\approx \lambda_{dB}/(2\pi\theta)$, where $\lambda_{dB}$ is the 
de Broglie wavelength of the colliding atom.
Therefore, when 
$
\theta > \sqrt{\lambda_{dB}/(2\pi d)},
$
a substantial number of atoms are lost by hitting the side wall.
For a periodicity of 100~$\mu$m this angle
is approximately 3~mRad.

\begin{figure}
\includegraphics[width=0.7\linewidth,height=0.5\linewidth]{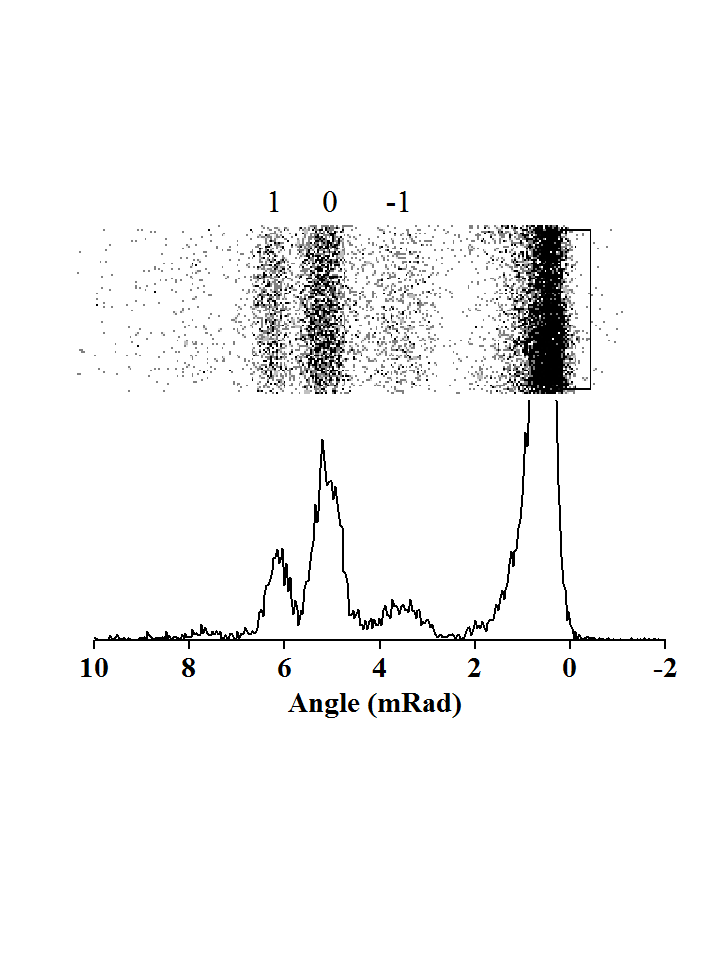}
\\
\includegraphics[width=0.7\linewidth,height=0.5\linewidth]{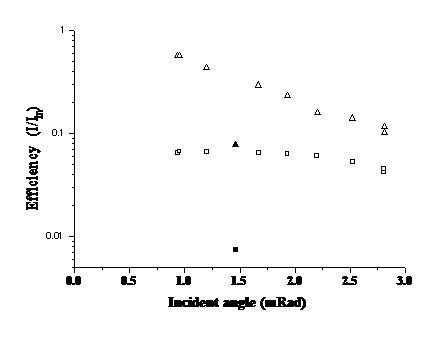}
\caption{(a) The diffraction pattern from a grating with
a pitch of 2~mm. The reflective part is composed of 
one hundred 100-nm-wide
ridges separated by 10~$\mu$m. (b) The efficiency vs the incident angle 
of the specular reflection (triangles) and 
the first-order diffraction (squares). The filled triangle and 
the filled square
are the efficiency of the specular reflection and the first-order
diffraction from a flat surface grating, respectively.}
\label{splitter}
\end{figure}

Further improvement of the reflectivity at a larger angle will be possible
if the width of the ridge and the periodicity are reduced. The reflecting
distance $z_{max}$ decreases as the effective 
$C_3$ decreases and also as $k_0$ increases. The ultimate limit on the
reflectivity is determined by the distance at which the colliding atom is
influenced by interactions other than van der Waals potential. When the
atom approaches too close to the surface, the interaction potential become
flattened by the repulsive core potential
and ceases to be sufficiently steep to cause reflection.
A metastable atom can be inelastically scattered when it touches 
the electron on the solid surface.
The distance to cause these phenomena
is in the order of nm. Therefore,
the ultimate normal incident velocity or the incident angle is determined by
$\lambda_{dB}/(2\pi\theta)\sim 10\mbox{nm}$. For the metastable
neon this gives the normal incident velocity of 50~cm/s. 
For a lighter atom such as helium
the reflectivity should be high at any angles in principle
if the atom is cooled below 100~$\mu$K.

As an example of application to atom optical elements, 
we fabricated a reflective
grating on a silicon surface with a ridged structure.
Figure~\ref{splitter} shows the diffraction grating 
with a periodicity of 2~mm.
The grating was composed of
1-mm-wide reflective stripes each of which was composed of $10^2$ ridges 
separated by 10~$\mu$m. The top of the ridge was 100~nm wide.
At a large diffraction angle, the intensity of 
the first-order diffraction relative to the specular reflection
was approximately 0.4,
which is in agreement with the intensity ratio of a transmission
grating with a 50\% opening. As the incident angle is reduced,
the relative intensity of the first-order diffraction becomes weaker,
and the efficiency becomes almost constant at a value of 6.5~\%.
This is due to the same physics that increases the reflectivity of
a ridged surface. 
As the angle decreases, all atoms become deflected even when the spacing
of two successive ridges is as large as 1~mm. Therefore, the 1-mm-wide
flat stripe does not function
as an absorbing surface, and the contrast of the amplitude grating is
lost.
A similar loss of contrast was observed on the
phase grating of an evanescent wave\cite{Cognet}.

In conclusion we have demonstrated that for light atoms the reflectivity
of quantum reflection can be sufficiently increased for use as practical
reflective atom optics elements, which enables us to design simple 
stable devices. The present result shows that the reflectivity changes
markedly depending on the surface condition. Therefore, the quantum 
reflection is a valuable tool 
for studying the characteristics of solid surfaces.

\section*{Acknowledgments}
This work was partly supported
by the Grants in Aid for Scientific
Research (11216202) from the Ministry of Education, Culture, Sports, 
Science and Technology.
One of the author(FS) is grateful to C. I. Westbrook for valuable 
discussions.


\begin{thebibliography}{99}


\bibitem{Shimizu} F. Shimizu: Phys. Rev. Lett. {\bf 86} (2001) 987,
and references therein.

\bibitem{Nayak} V. U. Nayak, D. O. Edwards, and N. Masuhara: Phys. Rev.
Lett. {\bf 50} (1983) 990.

\bibitem{Berkhout} J. J. Berkhout {\it et al}: 
Phys. Rev. Lett. {\bf 63} (1989) 1689.

\bibitem{Doyle} J. M. Doyle {\it et al}: Phys. Rev. Lett. {\bf 67}
(1991) 603.

\bibitem{Yu} I. A. Yu {\it et al}: Phys. Rev. Lett. {\bf 71} (1993) 1589.

\bibitem{Carraro} For more accurate discusions, see, for example,
C. Carraro and M. W. Cole: Prog. Surf. Sci. {\bf 57} (1998) 61.

\bibitem{Cognet} L. Cognet {\it et al}: Phys.Rev. Lett. {\bf 81} (1998) 5044.

\end{thebibliography}
\end{document}